\documentclass[a4paper,10pt,twoside]{cpc-hepnp}

\usepackage{multicol}
\usepackage{graphicx}
\usepackage{booktabs}
\usepackage{amssymb,bm,mathrsfs,bbm,amscd}
\usepackage[tbtags]{amsmath}
\usepackage{lastpage}

\begin{document}

\fancyhead[c]{\small  Submitted to ¡®Chinese Physics C'} \fancyfoot[C]{\small \thepage}

\footnotetext[0]{Received 20 October 2012}

\title{A study of pickup and signal processing for HLS-II bunch current measurement system \thanks{Supported by National Natural Science
Foundation of China (11105141) and the Fundamental Research Funds for the Central Universities (WK2310000015)}}

\author{%
      YANG Yong-Liang$^{1)}$\email{ylyang@ustc.edu.cn}%
\quad Ma Tian-Ji
\quad SUN Bao-Gen$^{2)}$\email{bgsun@ustc.edu.cn}%
\quad Wang Ji-Gang\\
\quad Zou Jun-Ying
\quad Cheng Chao-Cai
\quad Lu Ping
}
\maketitle

\address{%
National Synchrotron Radiation Laboratory (NSRL), University of Science and Technology of China, Hefei 230029, China\\
}

\begin{abstract}
For the HLS-II bunch current measurement system, in order to obtain the absolute value of bunch current, the calibration factor should be determined by using DCCT. At the HLS storage ring, the stretch effect of bunch length is observed and the change rate is about 19\% when the bunch current decays over time and this will affect the performance of bunch current detection. To overcome the bunch stretch influence in the HLS-II bunch current measurement, an evaluation about pickup type and signal processing is carried out. Strip-line pickup and button pickup are selectable, and the theoretical analysis and demonstration experiment are performed to find out an acceptable solution for the bunch current measurement system at HLS-II. The experimental data analysis shows that the normalized calibration factor will change by about 27\% when the bunch length change by about 19\% if using the button pickup and processing by peak value of bunch signal, the influence will be reduced to less 2\% if adopting the strip-line pickup and integral.
\end{abstract}

\begin{keyword}
bunch current, bunch length, strip-line electrode, button electrode, BPM, DCCT
\end{keyword}

\begin{pacs}
29.20.Dh,29.27.Bd,29.27.Fh
\end{pacs}

\footnotetext[0]{\hspace*{-3mm}\raisebox{0.3ex}{$\scriptstyle\copyright$}2013
Chinese Physical Society and the Institute of High Energy Physics
of the Chinese Academy of Sciences and the Institute
of Modern Physics of the Chinese Academy of Sciences and IOP Publishing Ltd}%

\begin{multicols}{2}

\section{Introduction}

HLS-II is a dedicated second generation VUV light source, and its storage ring operates in 800MeV with 204.016MHz RF and 45 bunches, the bunches are separated from each other by only 5 ns, and the bunch length is about 300ps. Bunch current is an important parameter for studying the injection fill-pattern in the storage ring and the instability threshold of the bunch. Especially, the bunch current monitor is a necessary tool for the top-up injection in an accelerator. A bunch by bunch current measurement (BCM) system\cite{lab1} has been developed to meet the needs of the upgrade project of Hefei Light Source (HLS-II).

Various types of bunch current or bunch charge measurement systems are developed world wide. Fast current transformers, electrodes-pickups and wall current monitors are the most popular signal pickups for bunch current measurement. Typically, a beam position monitor (BPM) has four electrodes, and the sum signal of the four electrodes carries the bunch charge information and its change rate is less than 0.005 when the beam position is alternated within 4mm\cite{lab2}. So the sum signal can be used to calculate the bunch current. At BEPC-II and SSRF, button electrodes BPMs are selected as signal pickups of bunch current detection\cite{lab3}\cite{lab4}.

Two types of four-electrodes pickup can be selected as the pickup for HLS-II bunch current measurement system: button electrode and strip-line electrode. The peak value or integral of bunch signal from pickup can be used to calculate the related bunch current value. To obtain the absolute value of bunch current, the calibration factor should be determined by using DC current transformer (DCCT). At HLS, the stretch effect of bunch length was observed\cite{lab5} when the bunch current decayed over time, and this will affect the performance of bunch current detection for different pickups and calculation methods, which have not been taken into consideration in other accelerators. So, to find out an ideal solution for the bunch current measurement at HLS-II, the evaluation about pickup type and signal processing is presented in this paper.

\section{The signal pickup and calculation technique}
Button electrode and strip-line electrode are candidate signal pickups for the HLS-II bunch current measurement. To overcome the bunch stretch influence on the bunch current detection, it is necessary to understand the pickup sum signal and to find the effective signal processing technique.

The electrons in a bunch in an electron storage ring are usually expressed with a Gaussian distribution\cite{lab1}. When the total charge in bunch is $Q_{0}$ and the bunch length is $\sigma_\tau$, Eq.(\ref{eq1}) shows the expression of a bunch in time domain.
\begin{equation}
\label{eq1}
I_b(t)=\frac{Q_0}{\sqrt{2\pi}\sigma_\tau}\exp(-\frac{t^2} {2\sigma_\tau^2}).
\end{equation}
For the button type pickup, the sum signal of four electrodes can be expressed as Eq.(\ref{eq2}):
\begin{equation}
\label{eq2}
V_{b\_\Sigma}(t)=-k\frac{dI_b(t)}{dt}=k\frac{Q_0}{\sqrt{2\pi}\sigma_\tau^2}t\exp(-\frac{t^2} {2\sigma_\tau^2}).
\end{equation}
$k$ is the scale factor of electronics. The chart is shown in Figure 1(a). The peak value or integral value of each bunch sum signal carries bunch charge information, which can be obtained from Eq.(\ref{eq2}):
\begin{equation}
\label{eq3}
V_{b\_peak}=K_p\frac{Q_0}{\sigma_\tau^2}\propto\frac{Q_0}{\sigma_\tau^2}
\end{equation}
\begin{equation}
\label{eq4}
V_{b\_integral}=\int_{t1}^{0}V_\sigma(t)dt=K_I\frac{Q_0}{\sigma_\tau}\propto\frac{Q_0}{\sigma_\tau}
\end{equation}
where $K_p$ is the calibration factor for using the peak value of sum signal and $K_p$ is the calibration factor for using the integral of sum signal. The above equation shows the influence of the bunch length $\sigma_\tau$, both on $V_{b\_peak}$  and $V_{b\_integral}$.

Alike, for the strip-line electrode, the sum signal can be expressed as Eq.(\ref{eq5}) and the chart is shown in Figure \ref{fig1}(b) :
\begin{equation}
\label{eq5}
V_{s\_\Sigma}(t)=-\frac{\varphi Z}{4\pi}[I_b(t)-I_b(t-\frac{2l}{c})].
\end{equation}
 From  Eq.(\ref{eq5}), $V_{peak}$  and $V_{integral}$ are approximated as follows:
\begin{equation}
\label{eq6}
V_{s\_peak}=K_p\frac{Q_0}{\sigma_\tau}\propto\frac{Q_0}{\sigma_\tau}.
\end{equation}
\begin{equation}
\label{eq7}
V_{s\_integral}=\int_{t1}^{t2}V_\sigma(t)dt=K_I{Q_0}\propto{Q_0}.
\end{equation}

 Eq.(\ref{eq6}) shows that the peak value of sum signal of strip-line electrodes $V_{s\_peak}$ is in proportion to the bunch length $\sigma_\tau$ and at the same time is in inverse proportion to ${Q_0}$. However, Eq.(\ref{eq7}) shows that the integral value $V_{s\_integral}$  is proportional to bunch charge $Q_0$ and has no influence on the bunch length $\sigma_\tau$.

\begin{center}
\includegraphics[width=8cm]{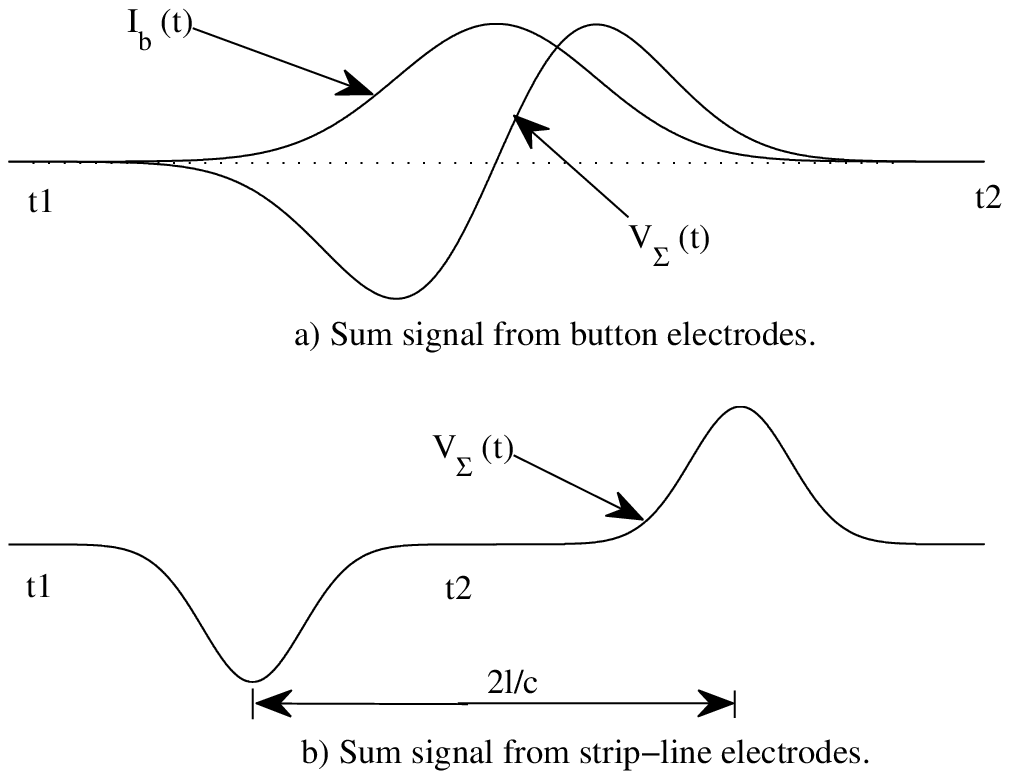}
\figcaption{\label{fig1}  The expression of a bunch in time domain and sum signal from BPM. }
\end{center}

\section{Experimental data analysis}
To evaluate and verify how the bunch length influences the calibration factor of bunch current measurement in different pickup types, the beam current (by DCCT), bunch length (by streak camera) and the bunch current are monitored at the same time in the HLS electron storage ring. The data acquisition unit for bunch current measurement is a digital oscilloscope (Agilent MSO7104). An 800MHz low pass filter is used to reduce the leak of frequency spectrum for the limited bandwidth of digital oscilloscope.

\subsection{Bunch length stretch effect at HLS}
Figure \ref{fig2} shows the DCCT beam current value decays over time. The beam current decays from 200mA to 80mA within about 10 hours when the HLS storage ring is in 800MeV energy and 45 bunches  operation mode.
\begin{center}
\includegraphics[width=8cm]{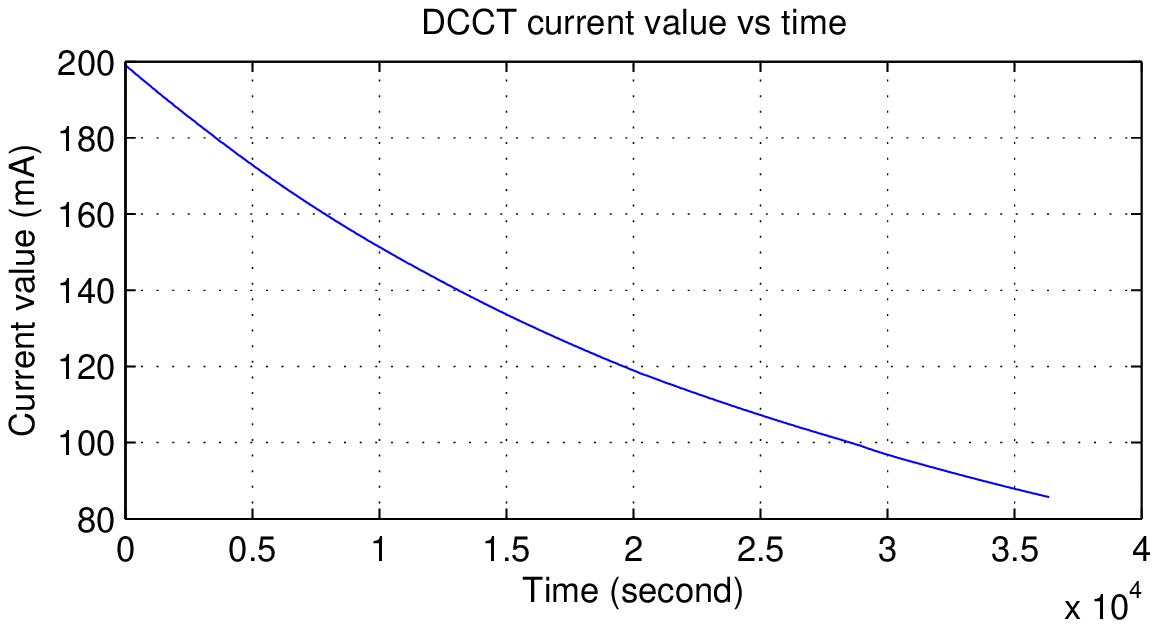}
\figcaption{\label{fig2}   The DCCT beam current value. }
\end{center}
At the same time, the bunch length can be obtained from streak camera. With the beam current decay\cite{lab5},
the stretch effect of bunch length is observed. Figure \ref{fig3} shows that the bunch length (averaged) changes from 303ps to 244ps with the beam current decay and the change rate is about 19\%.
\begin{center}
\includegraphics[width=8cm]{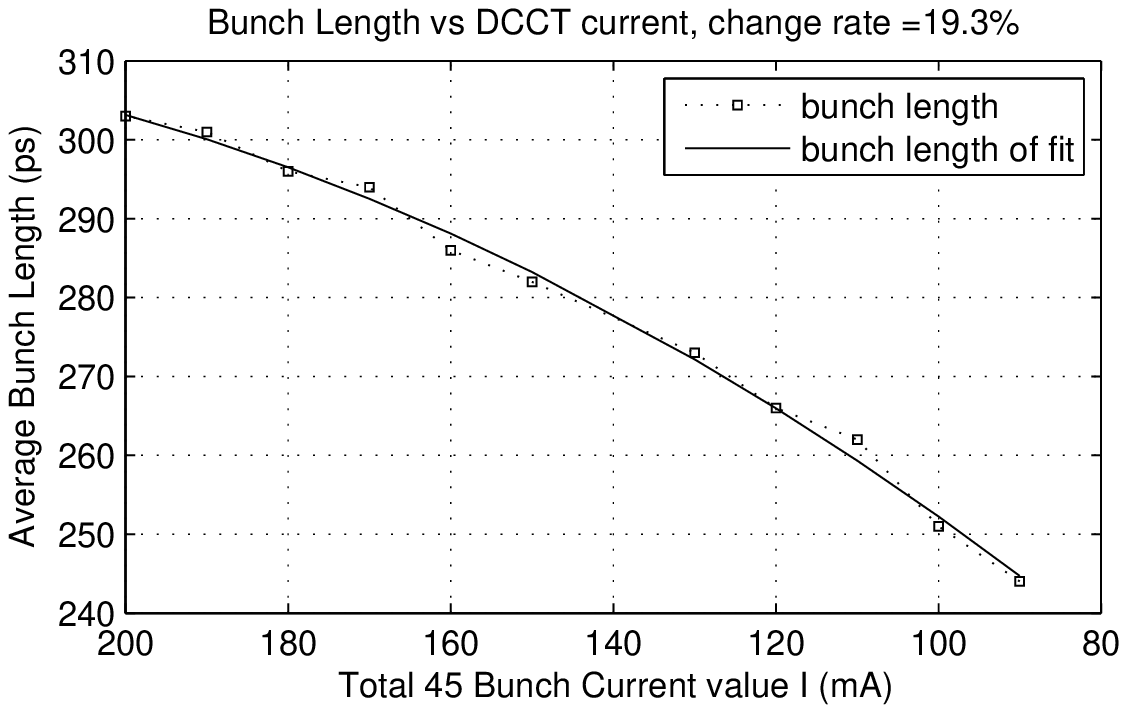}
\figcaption{\label{fig3}   Bunch length changed with bunch current. }
\end{center}

\subsection{Bunch Signal processing}
The RF of HLS is 204MHz, so the bunch spacing is about 5ns, and the width of bunch pulse after the extension by signal transmission network is about 300ps. The above parameters are same in HLS-II. The sampling rate of oscilloscope is 2GHz per channel and is not enough to obtain the integrate value or peak value of every bunch in one revolution time. The way to solve this problem is by software. At HLS electron storage ring the life time of beam is about 7 hours. For bunch current detection, bunch current decay can be omitted in hundred turns. A waveform-reconstruction algorithm, that is, equivalent sampling, is used to improve time resolution\cite{lab6}. The idea of this algorithm is that mapping
sample values of each circle to the corresponding time of the first circle with a sorting algorithm, making use of the relationship that radio frequency is not in direct portion to sampling rate.

For HLS bunch current measurement system, fifty turns signal was recorded in every processing period and were rebuild to one turn signal, and an effective sampling rate of about 100GHz was obtained.

Sum signal from button and strip-line pickup have been recorded by the digital oscilloscope in two sampling channel at the same time. Figure \ref{fig4} show the rebuild waveform of sum signal from four-strip-line pickup, include all 45 bunches.
\begin{center}
\includegraphics[width=8cm]{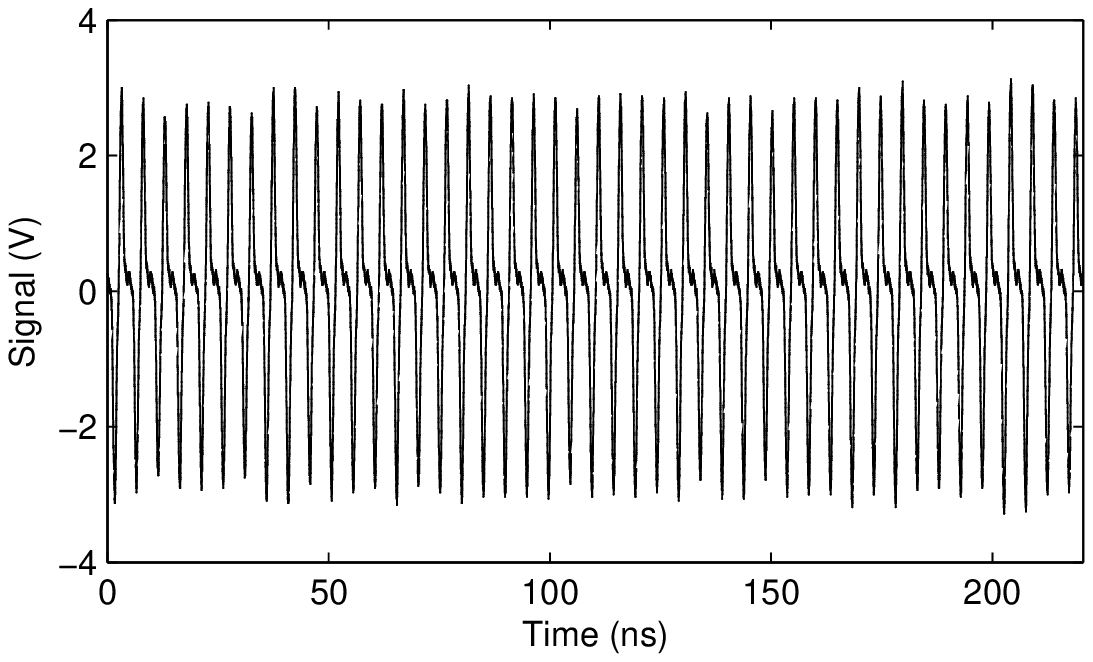}
\figcaption{\label{fig4} The sum signal of four-strip-line pickup with 45 bunches. }
\end{center}

Figure \ref{fig4-1} show waveform of one bunch from strip-line type pickup and Figure \ref{fig4-2} show waveform from button type pickup. Each bunch shape can be restored very well.
\begin{center}
\includegraphics[width=8cm]{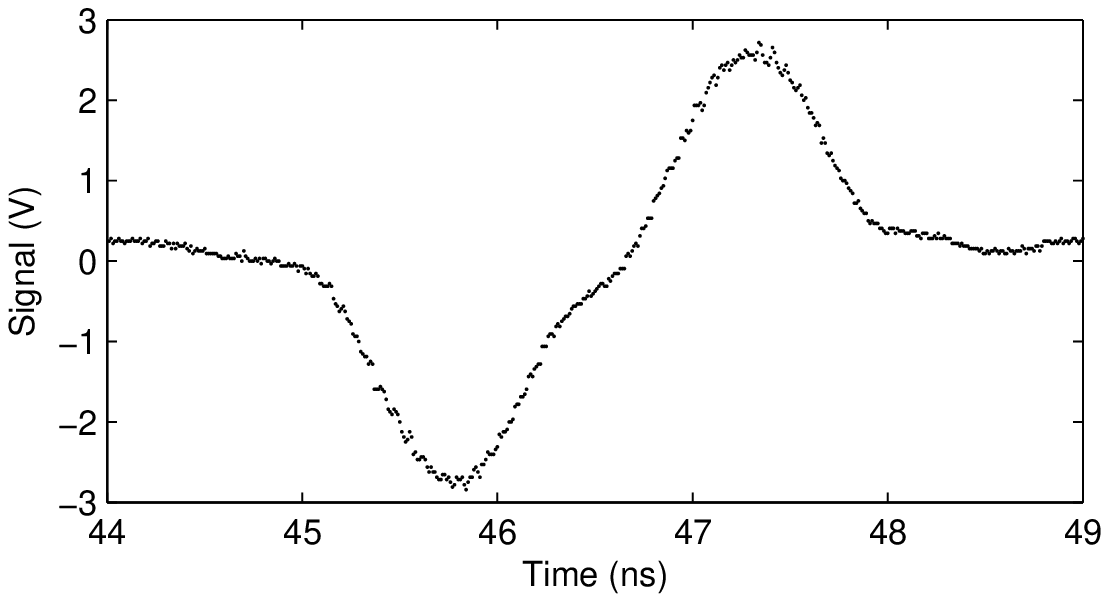}
\figcaption{\label{fig4-1} The sum signal of four-strip-line pickup with one bunch. }
\end{center}
\begin{center}
\includegraphics[width=8cm]{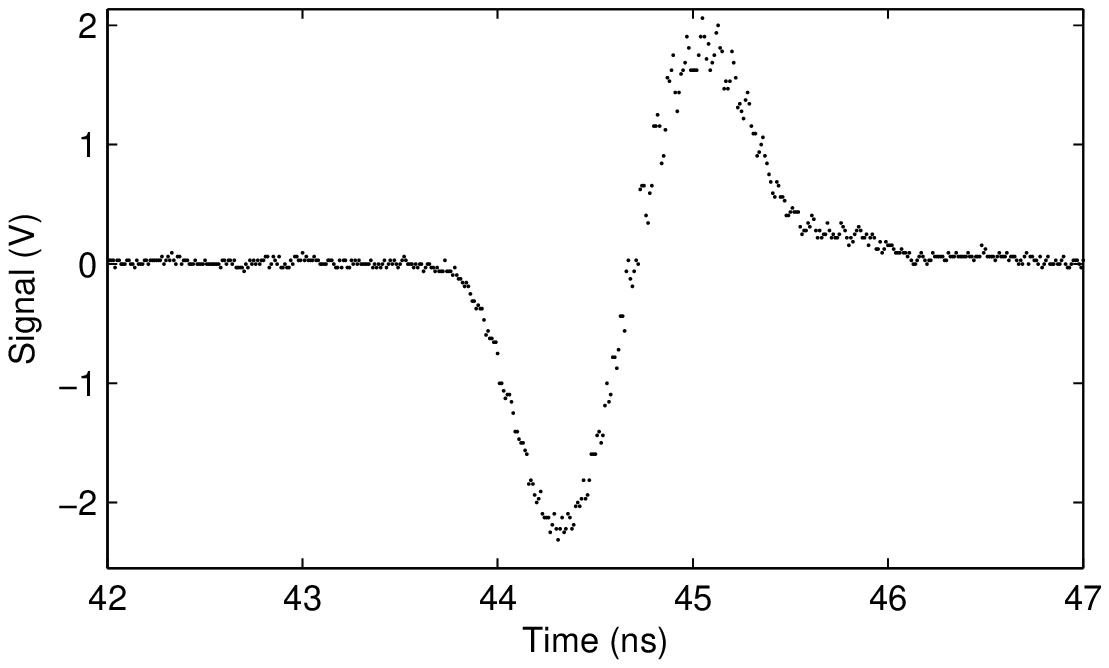}
\figcaption{\label{fig4-2} The sum signal of four-button pickup with one bunch. }
\end{center}

\subsection{Calibration factor for bunch current}
The amplitude of each bunch called the relative bunch current from the sum signal of BPM is calibrated, and normalized to get the absolute value of each bunch current. In the case of ignoring the difference of the calibration coefficient for each bunch current, each bunch current is calculated as follows:
\begin{equation}
\label{eq8}
I_i=KA_i,\qquad i=1,2,3,\dots,N
\end{equation}
where $A_i$ is the peak value or the integral value of the sum signal of BPM, $K$ is the calibration factor, $N$ is the number of bunches.
The calibration factor can be calculated with Eq.(\ref{eq9})
\begin{equation}
\label{eq9}
K=\frac{I_{dcct}} {\sum_{i=1}^{N} A_i} ,\qquad i=1,2,3,\dots,N
\end{equation}
where $I_{dcct}$ is the DC beam current obtained from the DCCT system.

With the beam current decay from 200mA to 80mA, we can calculate the calibration factor which changes following the beam current. The normalized calculation results are showed in Figure \ref{fig5}. With the strip-line pickup and integral (Fig.\ref{fig5} a), the normalized calibration factor changes only by 1.4\% when the bunch length changes by 19.3\%, the measurement RMS is 0.003. The factor change rate is 16.1\% with the button pickup and integral (Fig.\ref{fig5} b), 15.5\% with the strip-line pickup and peak value, and 27.1\% with the button pickup and peak value.

Figure \ref{fig6} shows the related normalized calibration factor error distribution and its RMS.
\end{multicols}

\begin{center}
\includegraphics[width=16cm]{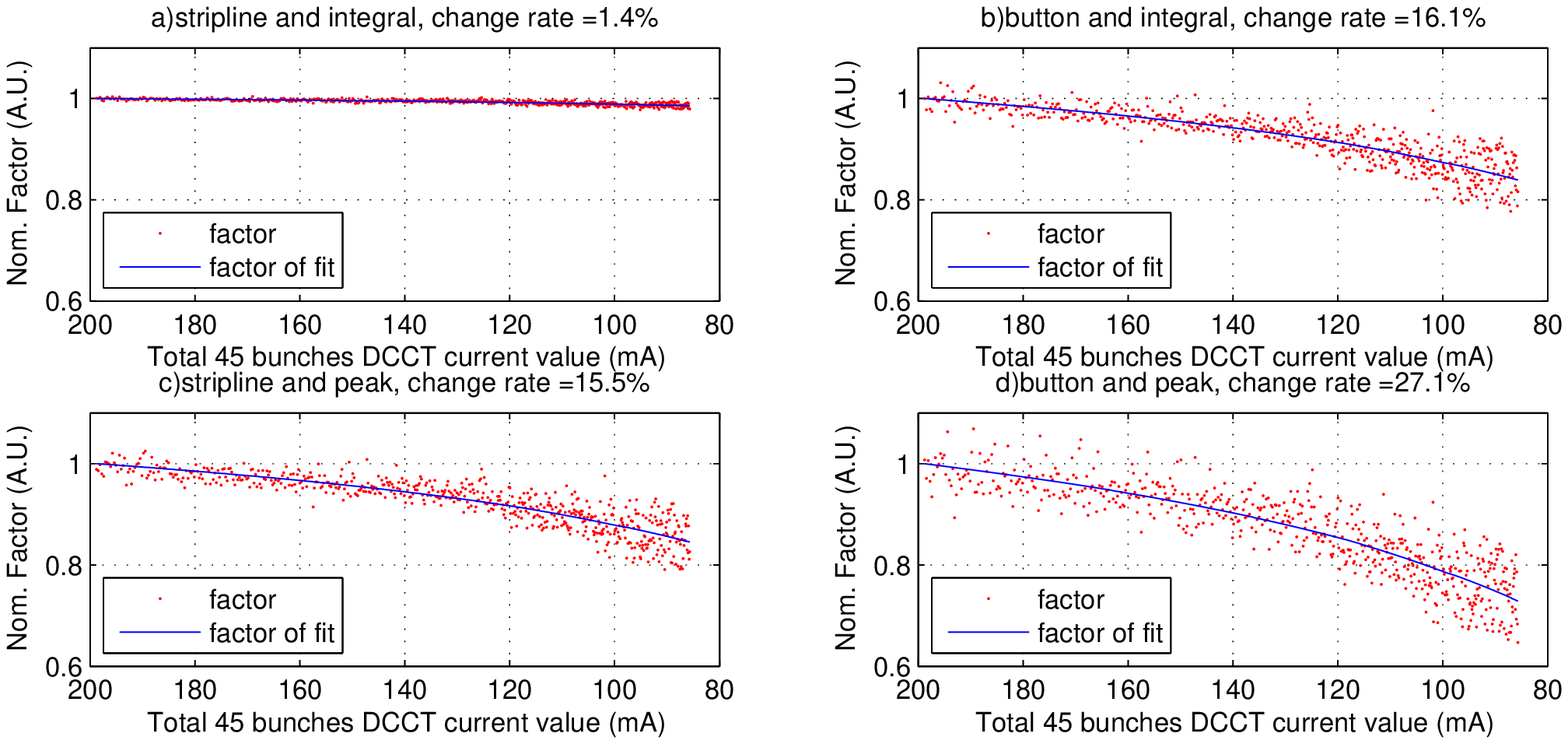}
\figcaption{\label{fig5} Calibration factor changing with bunch current value.}
\end{center}

\begin{center}
\includegraphics[width=16cm]{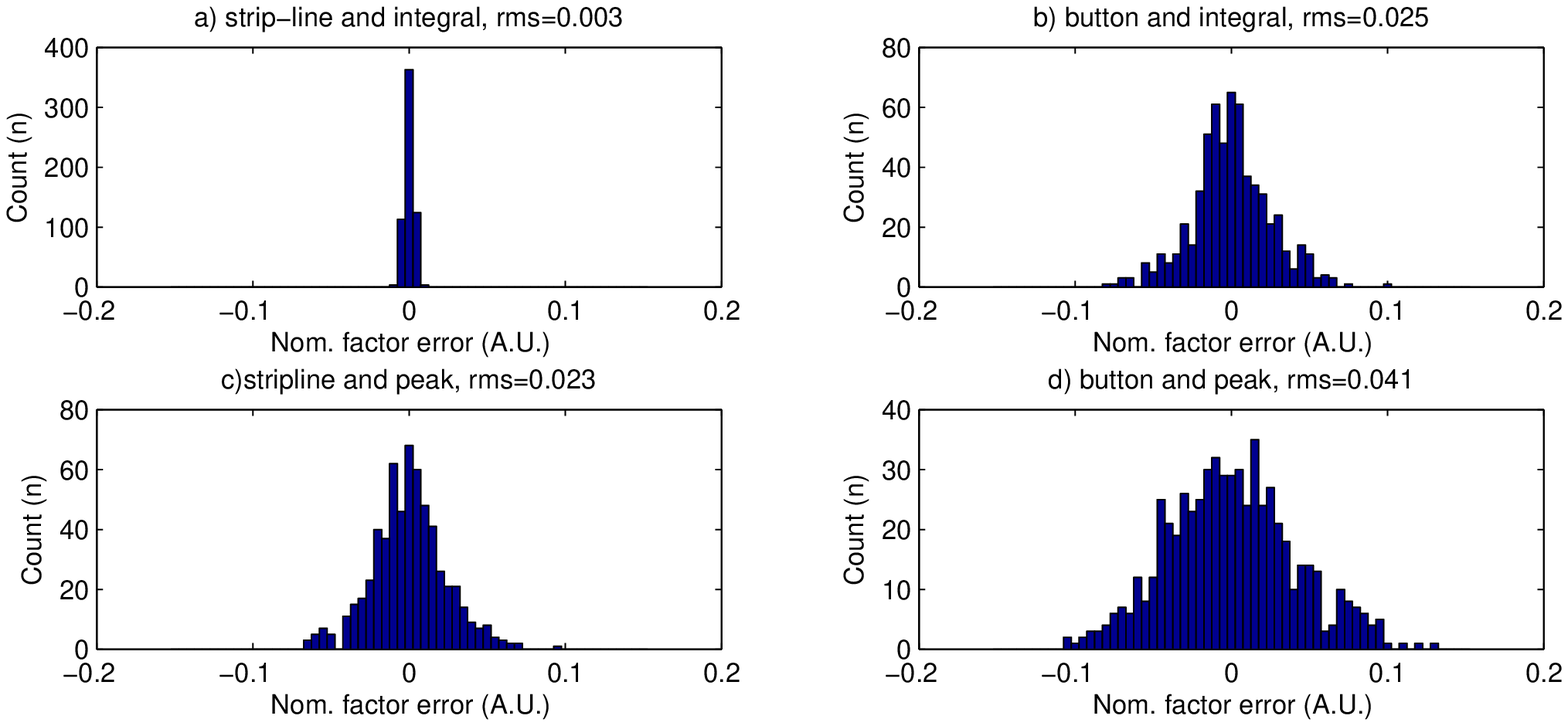}
\figcaption{\label{fig6} Normalized calibration factor error distribution.}
\end{center}

\begin{multicols}{2}

\section{Conclusion}

The theoretical and experiment data analyse show that using strip-line as signal pickup and processing the sum signal waveform by integral, the bunch length stretch effect on the calibration factor can be omitted. So, for the HLS-II bunch current measurement system, the strip-line pickup and integral method is finally selected to calculate the calibration factor and get the absolute bunch by bunch current value.\\

\acknowledgments{The authors would like to express their thanks to Prof. Weiming Li of NSRL for direct contributions of technical expertise. We also wish to thank Prof. Li Ma, Prof. Jianshe Cao and Dr. Jun-hui Yue (IHEP), and Prof. Yongbin Leng and Dr. Yingbing Yan (SSRF) for helpful discussions.}

\end{multicols}

\vspace{-1mm}
\centerline{\rule{80mm}{0.1pt}}
\vspace{2mm}

\begin{multicols}{2}

\end{multicols}

\clearpage

\end{document}